\documentclass[twocolumn,prd,showpacs]{revtex4}
\usepackage{graphicx}
\usepackage{amssymb}

\begin{document}

\title{\bf Inflation by spin and torsion in Poincar{\'e} gauge theory of gravity }
\author{S. Akhshabi}%
\email{s.akhshabi@gu.ac.ir}

\author{E. Qorani}%
\email{e.qorani@stu.gu.ac.ir}

\author{F. Khajenabi}
\email{f.khajenabi@gu.ac.ir}

\affiliation{Department of Physics, Golestan University, P. O. Box
49138-15759, Gorgan, IRAN} \vspace{1cm} 

\begin{abstract}
In Poincar{\'e} gauge theory of gravity, in addition to mass-energy
content, spin is also a source for gravitational interactions.
Although the effects of spin are negligible at low energies, they
can play a crucial role at very early universe when the spin density
was very high. In this paper by choosing a suitable Lagrangian for
Poincar{\'e} gauge theory of gravity, and suitable energy-momentum
and spin density tensors, we show that the effects of spin and
torsion can lead to a inflationary phase without the need for any
additional fields. No fine tuning of parameters is required in this
setup. We also calculate the scalar spectral index at the end of
inflation and show that it agrees with the most recent
observational data.\\

\end{abstract}

\maketitle
\section{Introduction}
Based on the pioneering works of Hermann Weyl, today, the standard
model of particle physics is described by a gauge theory of the
non-Abelian $SU(3)\times SU(2)\times U(1)$ group. The main idea of
gauge theories is that global symmetries are not compatible with
field theories and must be replaced with local ones. For keeping
physics invariant under local symmetries, we have to introduce some
compensating fields which then will describe the fundamental
interactions. This scheme applied to the symmetry group of standard
model, accurately describe three of the four fundamental
interactions, namely electromagnetic, weak and strong nuclear force.
The same framework can also be applied to gravitational interaction.
In the absence of gravity, the physical world is characterized by
the theory of special relativity which has the global Poincar{\'e}
transformations as its symmetry group. Localizing the Poincar{\'e}
transformations and demanding that the Lagrangian remains invariant
under the new local transformations, introduces two new fields which
turn out to be tetrad and spin connection fields. These new fields
(or similarly their field strengths, curvature and torsion tensors)
contain all the information about the gravitational interaction. The
resulting theory is called Poincar{\'e} gauge theory of gravity
(PGT), which contains general relativity as a special case. The
geometry of this theory is described by Riemann-Cartan space-time
which has both curvature and torsion \cite{BLAG}.

Throughout the paper, the Greek indices $\mu,\nu,..$ run over
$0,1,2,3$ and refer to the spacetime coordinates and Latin letters
$i,j,...$ run over $0,1,2,3$ and refer to the local Lorentz (or
tangent space) coordinates. In PGT tetrad and spin connection are
dynamical variables and fundamental geometric structures of the
theory.  The tetrad field is given as the components of  a set of 4
linearly independent vectors $e_{i}=e^{\mu}_{~i}\partial_{\mu}$
which form a basis in the tangent space on every point of the
manifold. The dual of this basis
$\vartheta^{i}=e_{~\mu}^{i}dx^{\mu}$ are coframes. Spin connection
$\Gamma^{~~j}_{\mu i}$ which is assumed to exist as an independent
field variable is related to the usual holonomic linear connection
$\Gamma^{~~~\rho}_{\mu\nu}$  by the relation
\begin{equation}
\Gamma^{~~j}_{\mu
i}=e_{~i}^{\nu}e^{~j}_{\rho}\Gamma^{~~~\rho}_{\mu\nu}+e_{~i}^{\nu}\partial_{\mu}e^{~j}_{\nu}
\end{equation}

The spacetime metric is not an independent dynamical variable here
and is related to the tetrad through the relations

\begin{equation}
g_{\mu\nu}=\eta_{ij}e_{~\mu}^{i}e_{~\nu}^{j},
\end{equation}
\begin{equation}
\textbf{e}_{i}\,.\,\textbf{e}_{j}=\eta_{ij}
\end{equation}
The inverse of the tetrad is  defined by
$e_{i}^{\mu}e_{\mu}^{j}=\delta_{i}^{j}$.

Torsion and curvature are given in terms of tetrad and spin
connection as
 \begin{equation}
  T_{\mu\nu}{}^{i} =2(\partial_{[\mu}e_{\nu]}^{~~i})
  +\Gamma_{[\mu|j }^{i} e_{|\nu]}^{j}),
 \end{equation}

 \begin{equation}
  R_{\mu\nu i}^{~~~~j} =2(\partial_{[\mu}\Gamma_{\nu]i}^{~~~j}
  +\Gamma_{[\mu|k}^{~~~~j} \Gamma_{|\nu]i}^{~~~~k}),
 \end{equation}
They also satisfy the  Bianchi identities
 \begin{eqnarray}
  &&\nabla_{[\mu}T_{\nu\rho]}^{~~~i} \equiv R_{[\mu\nu\rho]}^{~~~~~i},\\
  &&\nabla_{[\mu}R_{\nu\rho]}^{~~~ij}\equiv 0.
 \end{eqnarray}
General structure and physically acceptable Lagrangians of PGT have
been extensively studied in the literature, see for example
\cite{Hayashi, Sezgin, Rauch}. In \cite{Rauch} it has been argued
that in the class of theories described by $R+R^2$ Lagrangian, one
can evaluate the torsion in terms of the spin tensor. The general
Lagrangian of Poincar{\'e} gauge theory of gravity is a quadratic
function of curvature and torsion \cite{BLAG}
\begin{equation}
L_{G}=c_{4}R+L_{T}+L_{R}+c_{0}
\end{equation}
Here in this paper, we will only include even parity terms in the
Lagrangian, in this case we have
\begin{equation}
L_{T}=c_{1}T_{ijk}T^{ijk}+c_{2}T_{ijk}T^{jik}+c_{3}T_{i}T^{i}
\end{equation}
$$L_{R}=c_{5}R^{2}+c_{6}R_{ij}R^{ij}+c_{7}R_{ij}R^{ji}
+c_{8}R_{ijkl}R^{ijkl}$$
\begin{equation}
+c_{9}R_{ijkl}R^{klij}+c_{10}(\varepsilon_{ijkl}R^{ijkl})^{2}
\end{equation}
However cosmological implications of also including the parity
violating terms have been studied in \cite{Hehl,Ho}.  Field
equations are obtained by variation of Lagrangian with respect to
dynamical variables, tetrad field and spin connection (or
equivalently curvature and torsion). They take the form
\cite{Hayashi, Blag2, Shie}
\begin{eqnarray}
  \nabla_{\nu}H_{i}^{\mu\nu}-E_{i}^{~\mu}&=&{\cal T}_{i}^{~\mu},\\
  \nabla_{\nu}H_{ij}^{~~\mu\nu}-E_{ij}
  ^{~~\mu}&=&S_{ij}^{~~\mu},
 \end{eqnarray}
 where we have defined
 \begin{eqnarray}
  H_{i}^{~~\mu\nu}&:=&{\partial e L_{\rm G}\over \partial\partial_{\nu} e_{\mu}^i}
  =2{\partial e L_{\rm G}\over \partial T_{\nu\mu}{}^i},\\
  H_{ij}{}^{\mu\nu}&:=&{\partial e L_{\rm G}\over
   \partial\partial_{\nu}\Gamma_{\mu}^{~ij}}
  =2{\partial e L_g\over \partial R_{\nu\mu}{}^{ij}},
 \end{eqnarray}
 and
 \begin{eqnarray}
  E_{i} {}^{\mu}&:=&e^{\mu}{}_{i} e L_{\rm G}-T_{i \nu}{}^{j} H_{j} {}^{\nu\mu}
  -R_{i\nu}{}^{jk}H_{jk}{}^{\nu\mu},\\
  E_{ij}{}^{\mu}&:=&H_{[ij]}{}^{\mu}.
 \end{eqnarray}
The source terms here are energy-momentum and spin density tensors
respectively and are defined by
\begin{eqnarray}
 {\cal T}_{i}{}^{\mu}&:=&\frac{\partial eL_{\rm M}}{\partial e_{\mu}{}^i},\\
S_{ij}{}^{\mu}&:=&
    \frac{\partial eL_{\rm M}}{\partial \Gamma_{\mu}{}^{ij}},
\end{eqnarray}
where $L_{M}$ is the matter Lagrangian and $e$ is the determinant of
the tetrad.

 So there are two field equations in
Poincar{\'e} gauge theory of gravity and their sources are
energy-momentum and spin tensors. For a thorough review of gauge
theories of gravity including PGT and its cosmological solutions see
\cite{Blag2}.

\section{COSMOLOGY OF Poincar{\'e} GAUGE THEORY OF GRAVITY}
In this paper we consider a Lagrangian with the form
\begin{equation}
L_{g}=c_{1}T_{ijk}T^{ijk}+c_{2}T_{ijk}T^{jik}+c_{3}T_{i}T^{i}+c_{4}R+c_{5}R^{2}
\end{equation}
To find the field equations, first we should determine the suitable
form of energy-momentum, spin and torsion tensors by using
large-scale homogeneity and isotropy of the Universe (cosmological
principle). In Einestein-Cartan theory, the special case of
Poincar{\'e} gauge theory of gravity, Weyssenhoff fluid is usually
used to describe spin fluid \cite{Weys,Obuk}. However,  it can be
shown that the Weyssenhoff fluid description of the spin tensor is
not compatible with cosmological principle \cite{Tsam,BAUERLE,Goen}.
In cosmological solutions in Riemann-Cartan spacetime, where
connection is assumed to be independent of the metric, both the
metric and connection should satisfy the killing equation
\cite{Tsam}
\begin{equation}
L_{\xi}g_{\mu\nu}=0  \quad\quad    ;\quad\quad
L_{\xi}\Gamma_{~~\nu\rho}^{\mu}=0
\end{equation}
where $L$ is the Lie derivative in the direction of $\xi$. By the
above argument the only non-zero components of spin tensor are
\cite{Goen}
\begin{equation}
q(t)=S_{011}=S_{022}=S_{033}=-S_{i0i}
\end{equation}
\begin{equation}
s(t)=S_{123}=S_{312}=S_{231}=S_{[123]}
\end{equation}
Likewise, for torsion tensor
\begin{equation}
h(t)=T_{110}=T_{220}=T_{330}=-T_{i0i}
\end{equation}
\begin{equation}
f(t)=T_{123}=T_{312}=T_{231}=T_{[123]}
\end{equation}

Where due to cosmological principle,  $q$, $s$, $h$ and $f$ can only
depend on time. Moreover, similar to standard cosmology, we assume
that the energy-momentum tensor has the form of a perfect fluid.

The dual basis (or tetrad) is assumed to be homogenous and isotropic
FRW type in the form of
\begin{equation}
\vartheta^{0}=dt \quad ,\quad
\vartheta^{A}=a(t)(1+\frac{1}{4}kr^2)^{-1}dx^{A}
\end{equation}
$$
(k=0,\pm1) \quad ,\quad (A=1,2,3)$$ Using equation (2), this gives
the usual FRW metric
\begin{eqnarray}
 {\rm d}s^2 = -{\rm d}t^2+a^2(t)\Bigl[\frac{{\rm d}r^2}{1-kr^2}
 +r^2({\rm d}\theta^2+\sin^2\theta{\rm d}\phi^2)\Bigr]
\end{eqnarray}

By using the above definitions and equations (13-16) the field
equations are obtained by varying the Lagrangian with respect to
tetrad and spin connection

$$12c_{5}(M^{2}-N^{2})+\sigma(2hH-h^{2})-(c_{1}+3c_{2})f^{2}-2c_{4}N$$
\begin{equation}
=-\frac{1}{3}\rho
\end{equation}

$$12c_{5}(M^{2}-N^{2})+\sigma(3h^{2}-4hH-2\dot{h})+(c_{1}+3c_{2})f^{2}$$
\begin{equation}
+2c_{4}(2M+N)=-p
\end{equation}

\begin{equation}
2(2c_{4}-\sigma)h-24c_{5}(\dot{\varphi}-2h\varphi)=-q
\end{equation}

\begin{equation}
2(c_{4}-2c_{1}-6c_{2})f+24c_{5}f\varphi=-s
\end{equation}

Where we have defined

$$H=h+\frac{\dot{a}}{a}\quad\quad;\quad\quad M=\dot{H}+\frac{\dot{a}}{a}H$$
$$N=H^{2}+\frac{k}{a^{2}}-\frac{1}{4}f^{2}\quad\quad;\quad\quad F=\frac{1}{2}(\dot{f}+\frac{\dot{a}}{a}f)$$
$$\varphi=M+N=\frac{1}{6}R\quad;\quad \psi=fH+F=\frac{1}{12}R_{ijkl}\varepsilon^{ijkl}$$
$$\sigma=c_{1}+3c_{3}$$
Here a dot denotes differentiation with respect to time and $a(t)$
is the scale factor in FRW line element. The Bianchi identities (6)
and (7) have been used to simplify the above equations. The trace of
tetrad field equation is

$$(c_{1}+3c_{2}-\frac{1}{2}c_{4})f^{2}=\frac{1}{6}(\rho-3p)+(\sigma-2c_{4})(\dot{h}+h^{2}$$
\begin{equation}
+3\frac{\dot{a}}{a}h)-2c_{4}(\frac{\ddot{a}}{a}+(\frac{\dot{a}}{a})^{2}+\frac{k}{a^{2}})
\end{equation}

Furthermore, fluid equation describing the conservation of
energy-momentum and spin tensors is

\begin{equation}
\dot{\rho}+3\frac{\dot{a}}{a}(\rho+p)=3(sF-qM)
\end{equation}
Here we have used the definitions and notations of reference
\cite{Goen} (see also \cite{Kudin}). In addition to the above
equations, one should also specify a suitable form of the equation
of state $p=\omega\rho$. The first two of the field equations, e.g
equations (27) and (28) can be regarded as the PGT analogues to
Einstein field equations of general relativity and can be obtained
by variation of gravitational and matter Lagrangians with respect to
the tetrad field. They have components of the energy-momentum tensor
as their source. The next two field equations (29) and (30) however,
are unique to PGT and are obtained by variation with respect to the
spin connection field. The spin tensor components which are the
source of these two equations, can be regarded as negligible in
late-time cosmological dynamics due to very low density of spinning
matter. However, in the early universe, when the densities were
extremely high, they can play a crucial role in the dynamics. It
should be noted that the spin of particles in the early universe
were most probably randomly oriented and as a result will probably
cancel out when averaged over a macroscopic region. However, even if
this was the case, many of the corrections in the field equations
are quadratic in spin and will not vanish by averaging procedure. As
a result even in macroscopic limit, one should expect some new
dynamics due to effects of spin in the early universe
\cite{Kerlick,Gasp}. The field equations (27-30) together with the
equation of state give us five equations for seven unknown functions
$q$, $s$, $h$ , $f$ , $\rho$ , $p$ and the scale factor $a(t)$. To
proceed further in solving the equations, one need to specify an
exact model for describing the spin density tensor. Here one could
proceed in several ways. One way is to specify exact equation of
state parameter for the spin functions $q$ and $s$  in the form of
$q=\omega_{q}\rho$ and $s=\omega_{s}\rho$. This will bring down the
number of independent parameters to five and allow us to solve the
system of equations, however the choice of equation of state
parameters in this method is quite arbitrary and we will not employ
it here. Instead one could treat spin functions $q$ and $s$ as
macroscopically averaged quantities and perform the averaging
procedure to get a specific relation between $q$ , $s$ and $\rho$.
One assumes a universe filled with unpolarized particles of spin
$\frac{1}{2}$, then using the averaging procedure given in
references \cite{Gasp,Nurgaliev} we have
\begin{equation}
\sigma^2 =\frac{1}{2}\langle S_{ijk}S^{ijk} \rangle =
\frac{1}{8}\hbar^{2} A^{-2/(1+\omega)}_{\omega}\rho^{2/(1+\omega)}
\end{equation}
where $\omega$ is the equation of state of the perfect fluid and
$A_{\omega}$ is a dimensional constant depending on $\omega$.
Applying this relation to the spin tensor given by (21) and (22),
gives us the relation between $q$ , $s$ and $\rho$. Assuming that
$q$ and $s$ are macroscopically averaged quantities, equation (33)
gives
\begin{equation}
q^2+s^2= \frac{1}{48}\hbar^{2}
A^{-2/(1+\omega)}_{\omega}\rho^{2/(1+\omega)}
\end{equation}
One can also follow the method described in reference \cite{BAUERLE}
and describe the spinning particles quantum mechanically, i.e. with
a Dirac field. In this method the macroscopic average of the spin
density is obtained by the relativistic Wigner function formalism.
The macroscopic spin density tensor is given in this model by
\begin{equation}
S^{\mu\nu\rho}=\epsilon^{\mu\nu\rho\tau}S_{\tau}
\end{equation}

where $\epsilon^{\mu\nu\rho\tau}$ is four-dimensional Levi-Civita
tensor and $S_{\tau}$ is macroscopic spin vector field defined in
equation (3.25) in reference \cite{BAUERLE}. Applying equation (35)
to (21) and (22) and noting that the spin tensor defined in (35) is
a completely antisymmetric tensor, we conclude that the spin
function $q(t)$ should be zero in this model. In the next section we
solve the set of equations (27-30) together with the equation of
state and relation (34) to determine seven unknown functions. To
bring the number of equations to seven, we impose the condition
$M=N$, which ensures that the tetrad field equation leads to the
usual Einstein's equation of GR in the appropriate limit
\cite{Goen}.  Equations (31) and (32) can be used to simplify the
field equations.

We are interested in the dynamics of the early universe, specially
the inflation era. The analysis is done using the most general
assumptions.

\section{DYNAMICS OF EARLY UNIVERSE AND INFLATIONARY EPOCH}
Usually studying the role of spin in the dynamics of early universe
has been done in the framework of Einstein-Cartan theory using a
Weyssenhoff fluid to describe the spin density tensor
\cite{Gasp,Popl,Cappo,Damia,Bohmer}. It has been shown that by using
a suitable spin fluid in Einestein-Cartan theory, we get to an
inflationary epoch, however in this framework the fine tuning of
parameters is required \cite{Gasp}.  Also inflationary solutions in
PGT has been studied in \cite{Mink1} but with the assumption of
vanishing spin tensor. Here we show that in PGT using the Lagrangian
in the form of (11), the effects of spin and torsion cause the
Universe to enter an inflationary epoch with a wide range of
parameters. Also, using the assumption of vanishing spin tensor at
late times, it has been shown that Poincar{\'e} gauge theory of
gravity is also able to explain present time accelerated expansion
of the universe \cite{Shie,Hehl,Chen,Li,Mink2}.
\subsection{Numerical results}
To find scale factor and other parameters we set
$$k=0,\quad c_{1}=0.025, \quad c_{2}=0.05,$$
$$c_{3}=-0.1\quad c_{4}=0.5,\quad c_{5}=10^{-4}$$ This choice
corresponds to the so-called 'scalar torsion mode' which has spin
and parity $0^{+}$ \cite{Yo}. For describing density and pressure of
perfect fluid we use the equation of state of radiation-dominated
Universe $p=\frac{1}{3}\rho$ in the following numerical
demonstrations, however any physically acceptable equation of state
parameter leads to the same cosmological dynamics at the very early
universe. We also use equation (33-34) with $\quad A_{\omega}=1$ to
describe macroscopically averaged spin tensor parameters . The
condition $M=N$ leads to the following equation
\begin{equation}
5.25\,
\dot{h}+4.33\,\frac{\ddot{a}}{a}+2.33\,\frac{\dot{a}^{2}}{a^{2}}+11.75\,h\,\frac{\dot{a}}{a}+3.25\,h^{2}=0
\end{equation}
Then, the system of equations (27-30) can be manipulated to give the
following relation between the scale factor $a(t)$ and torsion
function $h(t)$
\begin{equation}
h=\frac{1}{2}\frac{{\ddot{a}a+\dot{a}^{2}}}{a\dot{a}}
\end{equation}
For numerical analysis of equations, we have used two sets of
initial values

$$a(0)=0.001,\quad \dot{a}(0)=1,\quad h(0)=1$$
 $$a(0)=0.001,\quad \dot{a}(0)=1,\quad h(0)=-1 $$
However as can be seen from the figures, both sets of initial values
will yield  almost similar results. Figures $(1)$  shows the
qualitative evolution of the scale factor $a(t)$ and density
function $\rho(t)$ versus the cosmic time.  As we see in figure $1$,
the effects of spin and torsion causes the universe to enter an
inflationary epoch where the scale factor grows quasi-exponentially
with a positive acceleration.  There is no need for any additional
scalar (or any other) field in this setup. The question of graceful
exit from inflation should be considered in subsequent studies,
however figures $(2)$ show that two torsion functions $h(t)$ and
$f(t)$ and also two spin tensor functions $q(t)$ and $s(t)$ all
approach zero at late times. Consequently, the effects of spin
tensor in the large-scale dynamics of the universe will only be
relevant at the very early times. However, one should note that due
to the fundamental difference between the field equations of PGT and
their general relativistic counterparts, torsion can also effects
the late-time dynamics of the universe, as has been shown in
\cite{Hehl,Shie,Chen,Li}
\begin{figure*}[thbp]
\begin{tabular}{rl}
\includegraphics[width=7cm]{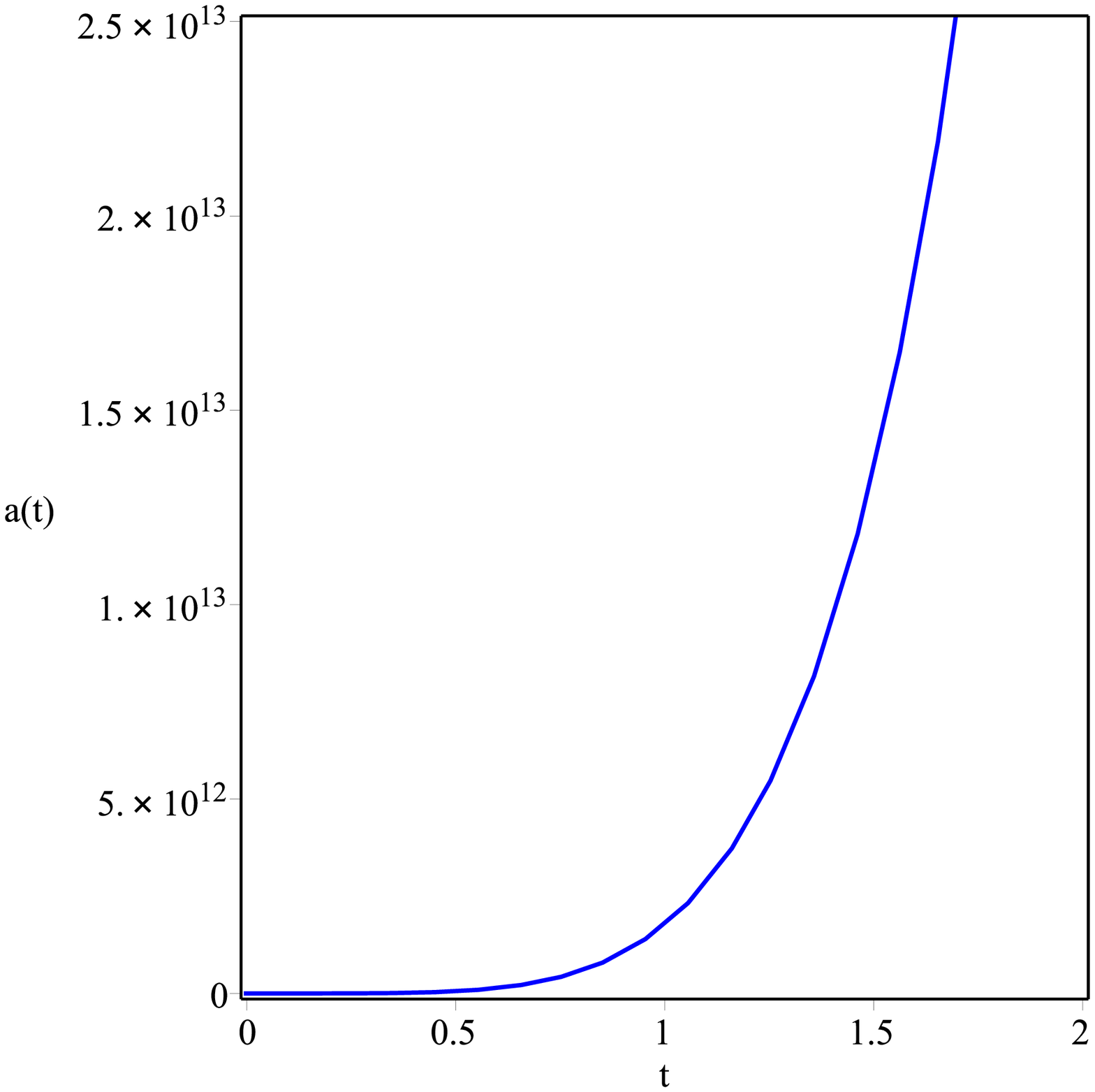}&
\includegraphics[,width=7cm]{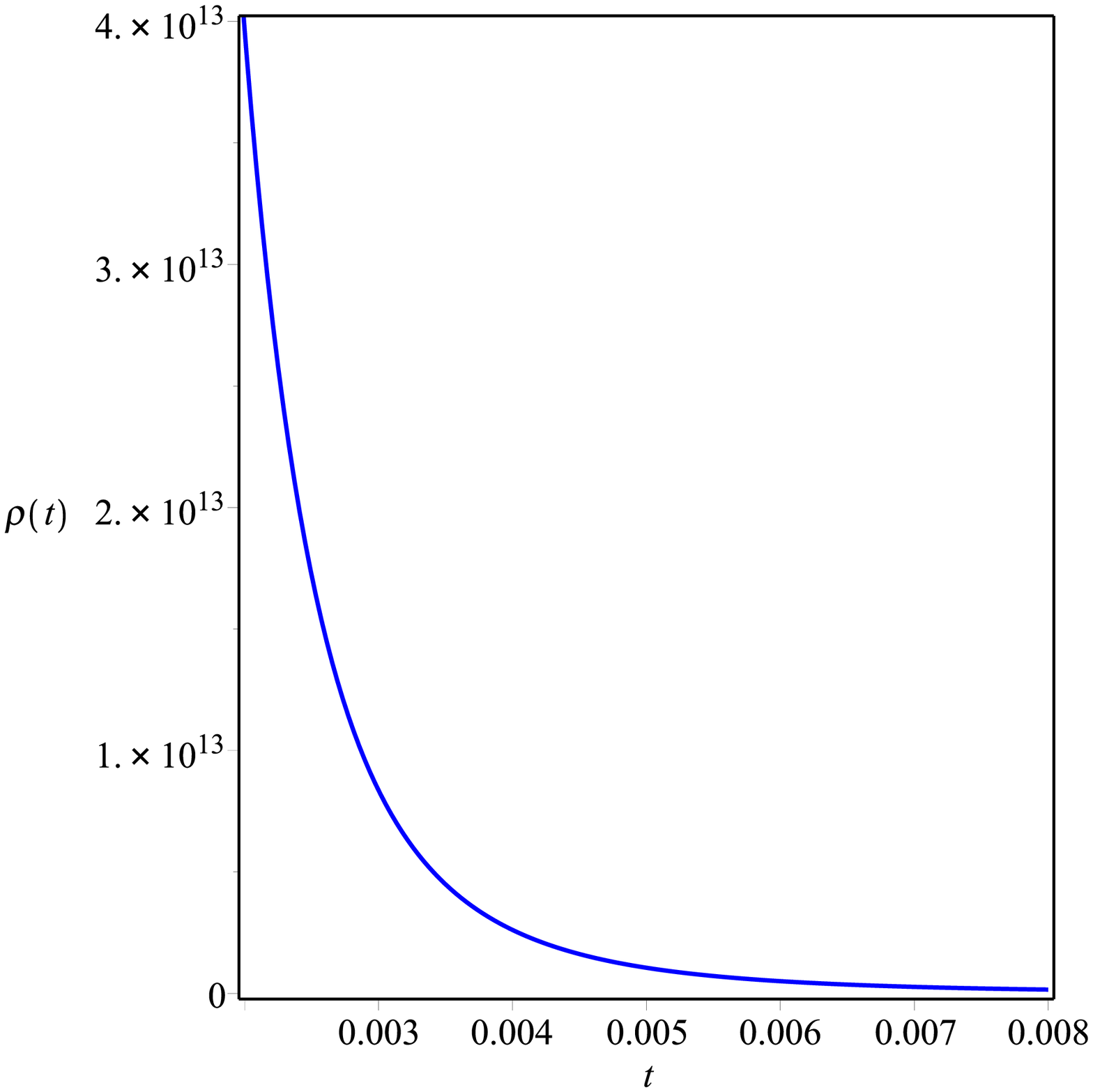}\\
\end{tabular}
\caption{\small{Evolution of scale factor $a(t)$ and the matter
density $\rho(t)$ as function of time.}}
\end{figure*}

\begin{figure*}[thbp]
\begin{tabular}{rl}

\includegraphics[width=7cm]{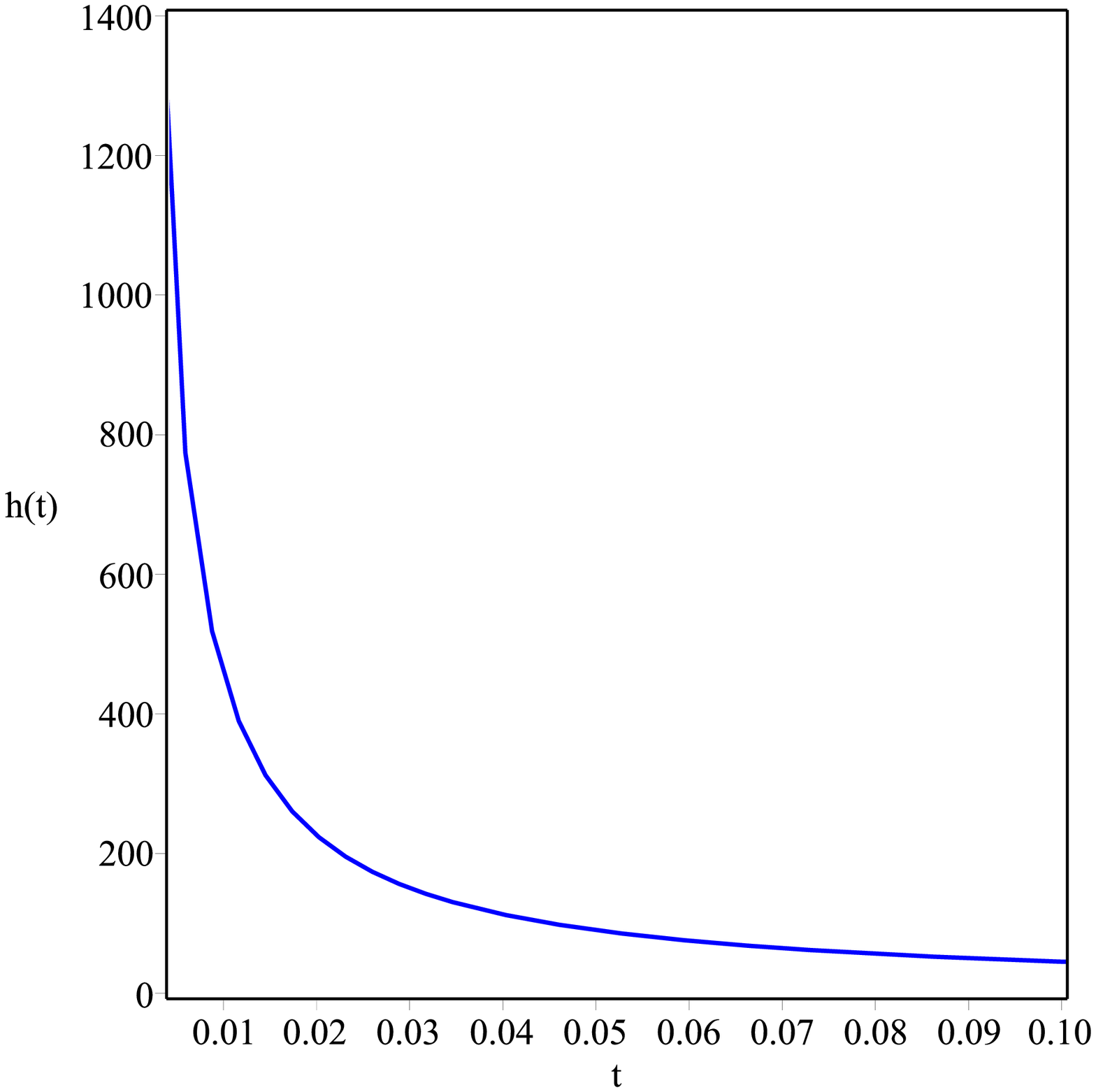}&
\includegraphics[width=7cm]{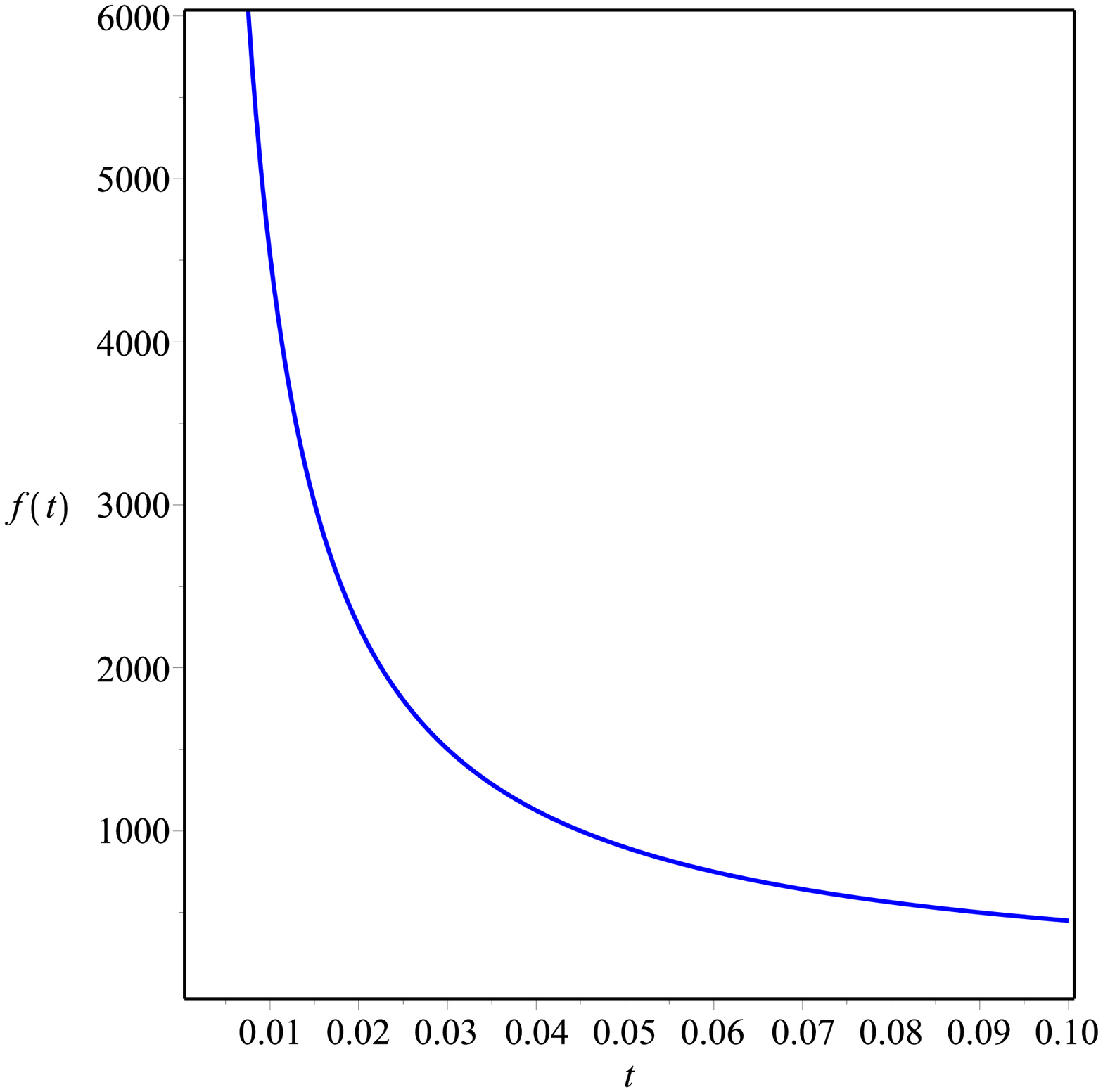}\\
\includegraphics[width=7cm]{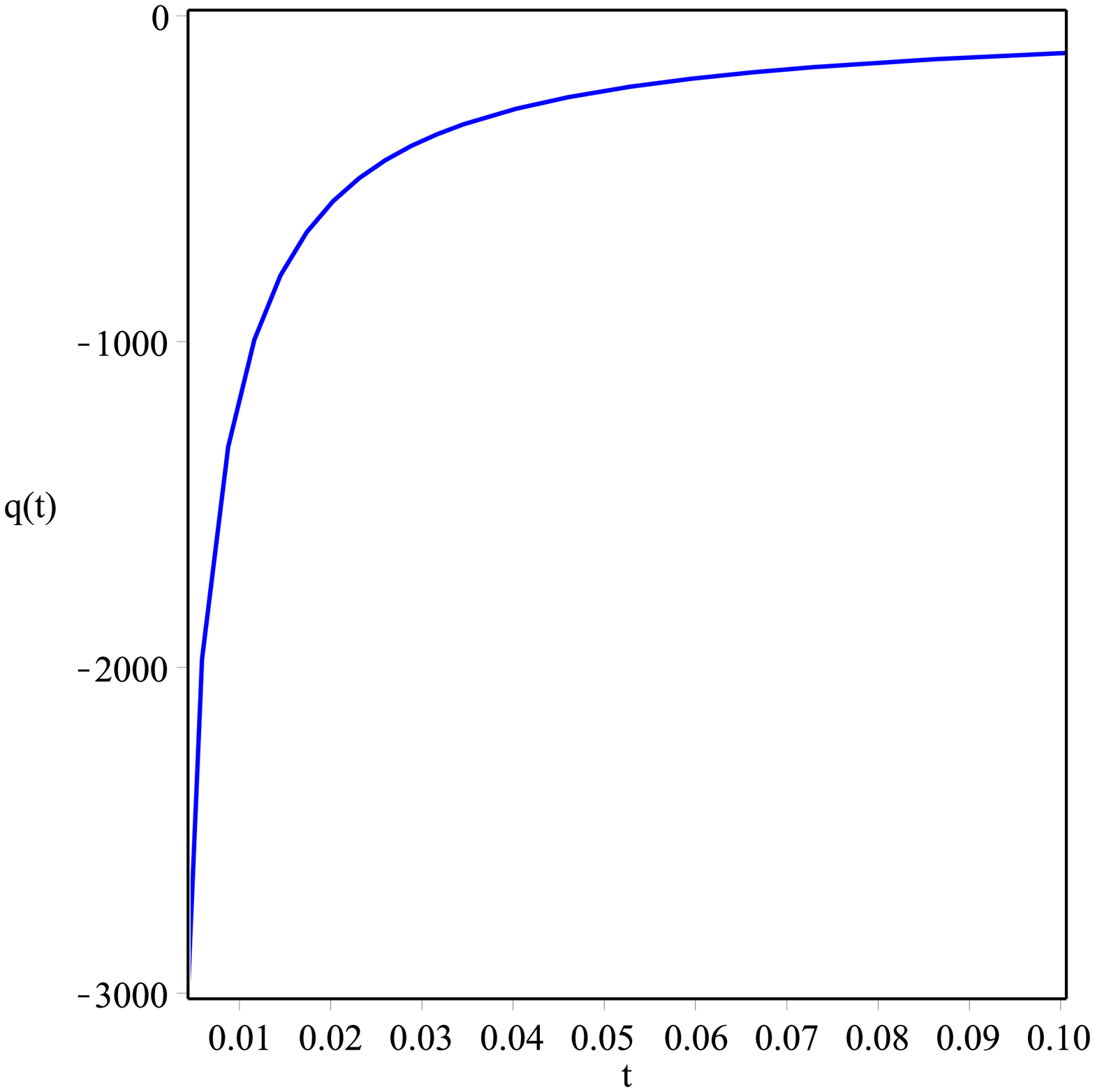}&
\includegraphics[width=7cm]{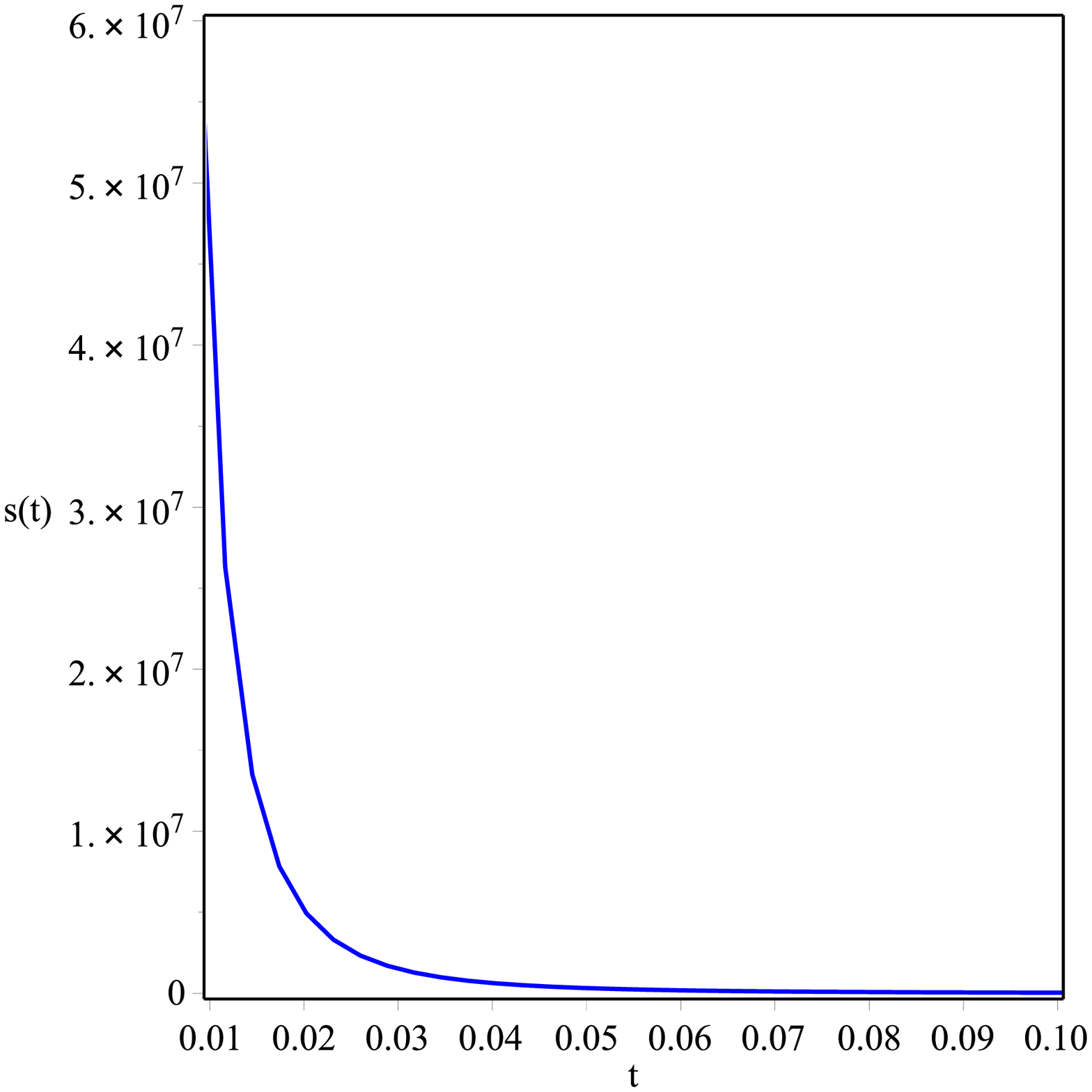}
\end{tabular}
\caption{\small{Evolution of torsion functions $h(t)$ and $f(t)$ and
spin density functions $q(t)$ and $s(t)$ as functions of time. They
all approach zero at late times as expected.}}
\end{figure*}
\begin{figure*}[thbp]
\begin{tabular}{rl}
\includegraphics[width=7cm]{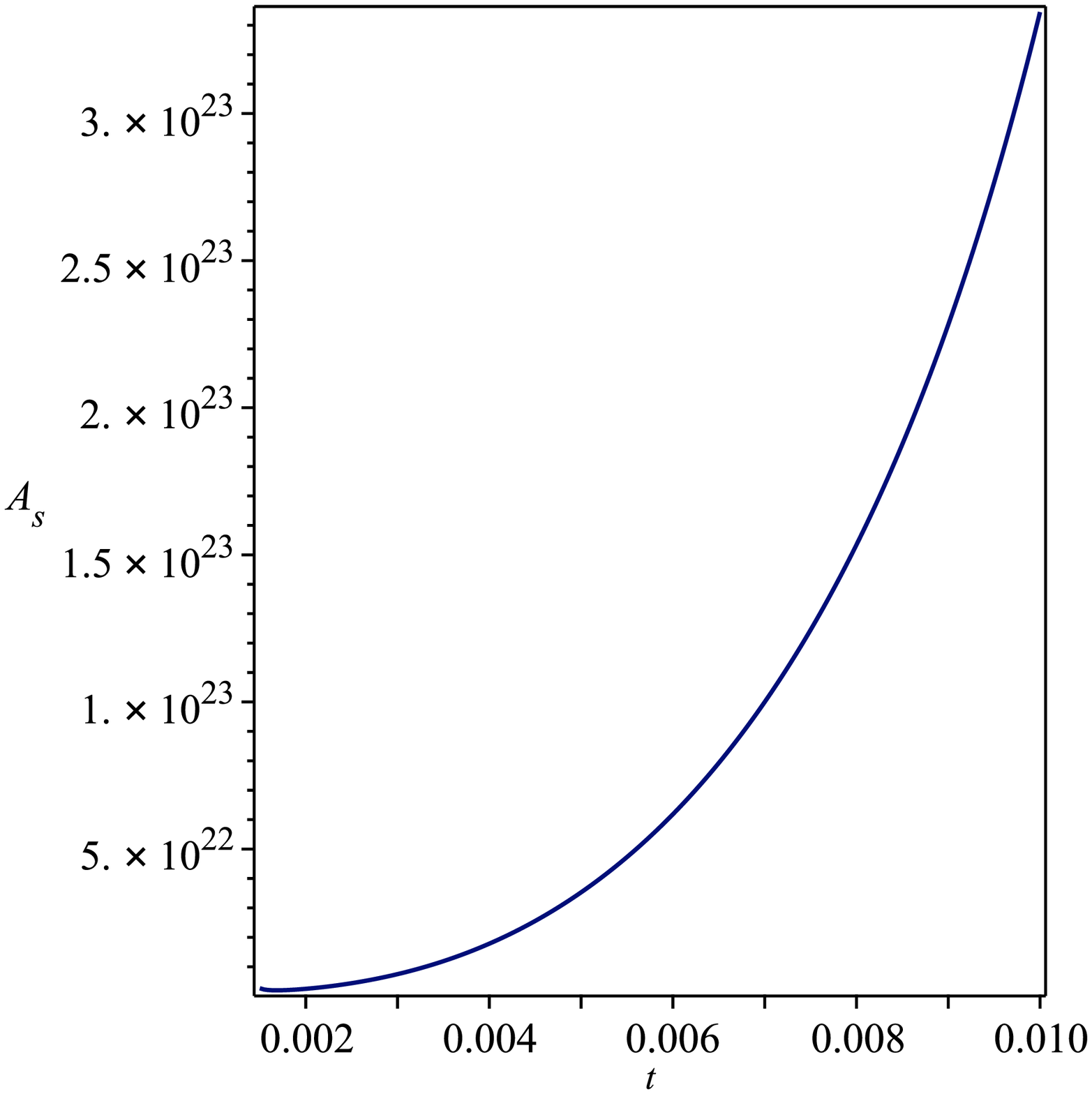}&
\includegraphics[width=7cm]{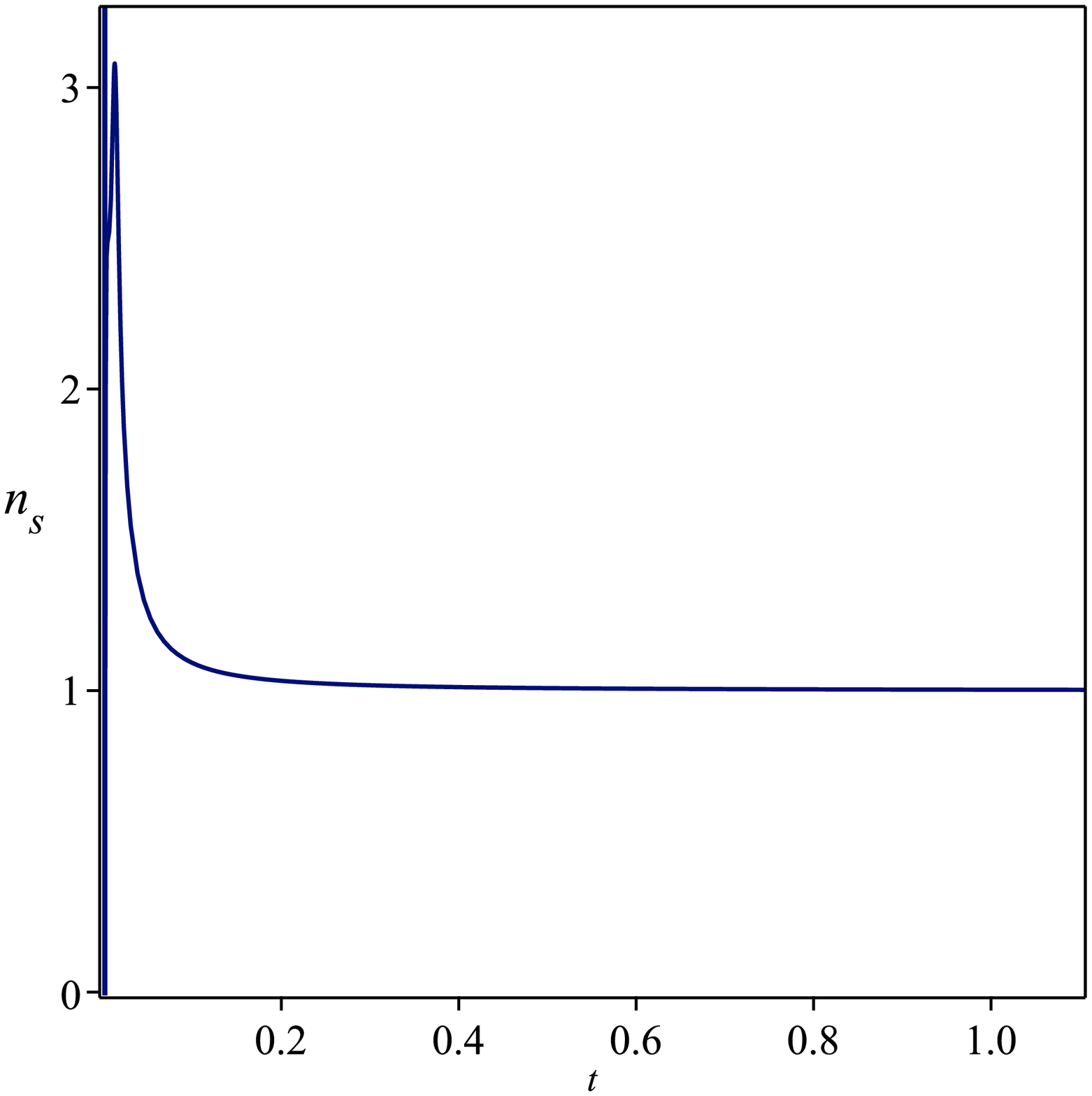}\\
\end{tabular}
\caption{The amplitude of scalar perturbations $A_{s}$ and The
scalar spectral index $n_{s}$ as function of time. $n_{s}$ is almost
1 at the time t=1.0001 which is the end of inflation as calculated
by equation (25).}
\end{figure*}

One of the most important tests for any inflationary model is its
ability to generate nearly scale-invariant spectrum of scalar
perturbations at the end of inflation. The scalar spectral index is
given by \cite{Lidd}

\begin{equation}
n_{s}(k)-1\equiv\frac{d\ln A_{s}^2}{d\ln k}
\end{equation}
where $A_{s}$ is the amplitude of scalar perturbations at the time
of horizon crossing
\begin{equation}
A_{s}(k)\equiv\frac{4}{5}\frac{H^{2}}{m_{pl}^{2}|H^{'}|}|_{k=aH}
\end{equation}
and
\begin{equation}
k=aH\exp[-N]
\end{equation}
N is the number of e-folding which can be calculated by
\begin{equation}
N\equiv\int H(t)dt
\end{equation}
Which we have obtained $N=35.11$ at the end of the inflation.
$A_{s}(k)$ and $n(k)$ are shown in figure (7) and (8).  As can be
seen in figure (3) the scalar spectral index approaches  unity at
the end of inflation which is compatible with observation that
predicts nearly scale-invariant but slightly red-tilted spectrum
\cite{Planck}.
\section{CONCLUSION}
Recently a cosmological model based on the Poincar{\'e} gauge theory
with quadratic Lagrangian has been constructed \cite{Shie}. This
model with suitable Lagrangian parameters, can explain the
accelerated expansion of the universe without the need for 'dark
energy'. By extending the work to the early universe, with a
non-zero spin density tensor, we numerically solved the set of field
equations of PGT and found the evolution of geometric and matter
parameters at the time of inflation. High density of particles with
spin can be considered as a source of torsion in early universe.
Although, because of random orientation of spin of particles, the
expectation value of spin tensor may be zero, nevertheless the
expectation value of the square of the spin tensor may not be so. As
a result, the effects of torsion play a crucial role in high
densities. In this paper by choosing a suitable Lagrangian for
Poincar{\'e} gauge theory of gravity and suitable spin and torsion
tensors, we have shown that the effects of spin and torsion can lead
to an inflationary epoch with exponential expansion in early
universe without the need for any additional fields. One of the most
fundamental test of any cosmological model is the issue of
cosmological perturbations. The scalar spectral index in this
framework also approaches the Harrison-Zel'dovich spectrum which is
perfectly consistent with recent observations. The number e-foldings
at the end of the inflation is also of the order of magnitude
required to solve the problems with standard big bang cosmology.
Here in PGT, in addition to perturbation modes originating from a
perturbed coframe, there are also independent connection modes. By
studying the dynamics and spectrum of extra modes, we can test the
validity of the model against current observational date and compare
the results to the standard models. For example, the extra modes
generating from a propagating spin connection, may result in some
identifiable signature in the anisotropy and non-Gaussianity
spectrum of the cosmic microwave background radiation.

\end{document}